\documentclass[english,pra,showpacs,twocolumn]{revtex4-1}
\usepackage[T1]{fontenc}
\usepackage[latin9]{inputenc}
\usepackage{color}
\usepackage{textcomp}
\usepackage{amsmath}
\usepackage{amssymb}
\usepackage{graphicx}

\makeatletter
\usepackage[colorlinks = true,
            linkcolor =blue,
            urlcolor  =blue,
            citecolor = blue,
            anchorcolor = blue]{hyperref}

\makeatother

\usepackage{babel}
\begin{document}
\title{Single-photon-added coherent state based measurement transition and
its advantages in precision measurement}
\author{Yusuf Turek}
\email{yusuftu1984@hotmail.com}

\author{Akbar Islam}
\author{Ahmad Abliz}
\email{aahmad@126.com}

\affiliation{School of Physics and Electronic Engineering, Xinjiang Normal University,
Urumqi, Xinjiang 830054, China}
\date{\today}
\begin{abstract}
In this work, the measurement transition as well as precision measurement
advantages of single-photon-added coherent state after postselected
von Neumann measurement are investigated. We noticed that the weak-to-strong
measurement transition characterized by the shifts of pointer's position
and momentum variables occurred in continuously by controlling a dimensionless
parameter associated with system-pointer coupling. We calculate the
ratio between the signal-to-noise ratios of nonpostselected and postselected
measurements, and the latter is used to find the quantum Fisher information.
We found that the single-photon-added coherent pointer state can improve
the precision of the measurement processes such as signal-to-noise
ratio and parameter estimation after postselected von Neumann measurement
characterized by postselection and weak value. Furthermore, contrary
to the results of several previous studies, we found that the anomalous
large weak values can't improve the precision measurement processes
related to single-photon-added coherent state. 
\end{abstract}
\pacs{03.65.-w, 03.65.Ta, 42.50.\textminus p, 03.67.\textminus a}
\maketitle

\section{introduction}

As we know, although the theory of quantum mechanics and its applications
are extremely successful, but its fundamental problems associated
with the measurement are still unsolved puzzles that the scientists
must to face. Handling the measurement problem requires to understand
the unique role of measurement in quantum mechanics. As a foundation
of quantum theory the quantum measurement problem were elaborated
in the 1930s by Bohr, Shrodinger, Heisenberg, von Neumann and Mandelstam.
Especially, in 1932 the von Neumann formulated \citep{Edwards1955}
the first standard version of quantum measurement characterized by
strong projective operation to the measured system so that it can
cause the wave packet collapse. 

According to the von Neumann's model, we can get the desired information
of the system observable by single trial since the coupling between
the measured system and pointer or meter is strong enough to distinguish
its different eigenvalues. This strong quantum measurement model promotes
the exploring of the quantum theory from its fundamentals \citep{Li:19,PhysRevLett.116.070404,RevModPhys.75.715,Wheeler1983,Khalili1992,Jacobs2014}
to quantum art technologies \citep{536,508,5361,417,549,521,351,9}.
However, because of the collapse of wave function owing to the decoherence,
the conventional strong quantum measurements cannot be directly used
in some hot topics of quantum information science such as quantum
state based high precision measurements \citep{PhysRevLett.125.080501,125,PhysRevLett.105.010405,PhysRevLett.111.033604},
reconstruction of unknown quantum state \citep{123,PhysRevA.83.032110,124,PhysRevA.63.040303},
etc. On the other hand, this strong measurement model becomes useless
if the coupling between measured system and pointer is too weak to
distinguish the different eigenvalues of the observable. In the latter
case, the sub-wave-packets corresponding to different eigenvalues
of system observable are overlapped so that we can't get the enough
information about the system beacause our subsequent trial may cause
the wave-packet's collapse \citep{PhysRevD.24.1516} . To solve this
issue, in 1988 Aharonov and co-workers proposed a generalized quantum
measurement technique called weak measurement which is characterized
by postselection and weak value \citep{PhysRevLett.60.1351}. In this
measurement model, even though the coupling strength is too small
compared to traditional strong measurement but still can get the required
system information statistically as accurate as strong projective
measurement by taking many measurement trials since it doesn't cause
wave function's collapse after measurement \citep{Tollaksen_2010}.
This new measurement model solved plenty of fundamental problems in
quantum theory and is widely accepted as a new measurement technique
(see \citep{RevModPhys.86.307} and references therein). However,
one may ask some questions like what is the connection between weak
and strong measurement? How can the weak-to-strong measurement transition
be characterized? In recent years, the above questions raised wide
interests among the physicists.

In general, the measurement results can be readout from the final
state of the pointer. In other words, the information of system's
observable in both strong and weak measurement schemes can be obtained
in the shifts of pointer's position and momentum after measurement.
Thus, the problems related to weak-to-strong measurement transition
could be explained by investigating the average values of the observable
showed in pointer's final shifts. To study the measurement transitions
in weak-to-strong regimes, we have to take into account the full order
evolution of unitary operator of interaction Hamiltonian \citep{Turek_2015,PhysRevA.85.012113}.
Zhu et al. \citep{PhysRevA.84.052111} studied the quantum measurement
transition for a qubit system without any restriction on the coupling
strength between the system and the pointer, and they found the transitions
from the weakest to strongest regimes. In a recent work \citep{Nature2020},
the weak-to-strong measurement transition is demonstrated experimentally
by using a single trapped $^{40}Ca^{+}$ ion. They found that the
weak-to-strong measurement transition predicts the continuous relation
between the weak value and expectation value of system's observable
showed in pointer's position shift, and the transition can be controlled
by modulating the coupling parameter $\Gamma=g/\sigma$. In this transition
factor, $g$ is the coupling constant between the measured system
and pointer and $\text{\ensuremath{\sigma}}$ is the width of an initial
Gaussian pointer, which characterizes the measurement strengths, i.e.,
$\Gamma<1$ ($\Gamma>1$) represents weak measurement (strong measurement
). In the above two theoretical and experimental works both considered
the fundamental Gaussian wave packet as a pointer. One of the authors
of the current work \citep{PhysRevA.92.022109} studied the average
shifts of pointer's position and momentum for a system observable
satisfying the property $\hat{A}^{2}=1$ by considering the full order
evolution of unitary operator of nonclassical pointer states including
squeezed vacuum state and Schrodinger cat state. They predicted that
the two extreme limits fit well with the weak and strong measurement
correspondences. Most recently, Orszag et al. \citep{PhysRevA.103.052215}
examined the weak-to-strong measurement transition for coherent squeezed
state, and theoretically proposed its implementation in $^{40}Ca^{+}$
ion stored inside the Paul trap interacting with two laser fields
as specific frequencies. They showed that the shift in the pointer\textquoteright s
position and momentum variables establishes a relationship with a
new value defined as the transition value, which can generalize the
weak value (weak measurement regime) as well as the conditional expectation
value (strong measurement regime). They extend the Josa's theorem
\citep{PhysRevA.76.044103}, and found that weak-to-strong measurement
transition can occur in continuously by modulating the coupling parameter
$\Gamma$. In \citep{Nature2020}, they used the fundamental Gaussian
mode as a pointer that correspond to the vacuum state case of the
measuring device in \citep{PhysRevA.103.052215}. As we know, the
fundamental Gaussian mode is the special case of coherent state and
its preparation is experimentally well established. The coherent state
is a semi-classical one with a Gaussian Wigner function, and if we
add one photon to the coherent state, one can obtain the non-Gaussian
quantum state called single-photon-added coherent (SPAC) state which
possesses nonclassical properties \citep{PhysRevA.43.492,Zavatta2004Quantum,PhysRevA.72.023820}.
As an intermediate state, the SPAC state covers both the single photon
state and the coherent state, which have potential applications in
quantum metrology \citep{Nph,PhysRevLett.105.113602,PhysRevLett.107.083601,PhysRevA.86.043828,HU201468,131,2015,Liu_2016,PhysRevA.98.012129,PhysRevLett.121.060506,PhysRevA.104.053712,Yuesf,1111}.
Furthermore, since the SPAC state has no vacuum component and contains
large single-photon probability, it bears more profound applications
in many fields including quantum state engineering \citep{129}, quantum
communication \citep{126}, quantum key distribution \citep{WANG20171393,RN1920,127,128,PhysRevA.90.062315},
and quantum metrology \citep{130,gard2016photon,PhysRevA.90.013821,SCHNABEL20171}.
Especially, in a recent work \citep{RN1921} the authors showed that
the performance of quantum digital signature can be improved significantly
by using the SPAC state compared with the one using weak coherent
state. Thus, the SPAC state may be a promising candidate for the implementation
of digital signature in the near future. 

In this paper, motivated by aforementioned works, we study weak-to-strong
measurement transition for the SPAC state. We also investigate the
advantages of SPAC state in some precision measurement processes such
as signal-to-noise ratio (SNR) and unknown parameter estimation after
postselected von Nuemann measurement. To achieve this, we take the
spatial and internal degrees of freedom of SPAC state as pointer and
measured system, respectively. We introduce the transition value of
system observable, and obtain the final normalized state of SPAC state
after postselection. We find that the weak-to-strong measurement transition
characterized by the shifts of pointer's position and momentum variables
occurred in continuously by controlling a dimensionless parameter
$\Gamma$ associated with system-pointer coupling. We also assertain
that the SPAC state based postselection measurement can improve the
SNR and Fisher information significantly for moderate weak values
and coupling parameter $\Gamma$. We anticipate that the results of
this work may help the further development of theoretical and experimental
schemes of weak-to-strong measurement transition based on SPAC pointer
state and their using in the above mentioned quantum information processing
and quantum metrology.

The rest of this paper is organized as follows. In Sec. \ref{sec:2},
we give the final pointer state after postseleceted measurement and
introduce some concepts related to the weak-to-strong measurement
transition. In Sec. \ref{sec:3}, we give the details of weak-to-strong
measurement transition controlled by coupling parameter $\Gamma$
between measured system and pointer. Subsequently, in Sec. \ref{sec:4}
and Sec. \ref{sec:5}, we investigate the advantages of SPAC state
in improving the efficiency of SNR and parameter estimations processes.
Finally, we summarize our findings of this study in Sec. \ref{sec:6}.

\section{\label{sec:2} Basic Concepts }

In quantum measurement theory, the related discussions usually begin
with the interaction Hamiltonian of the system and pointer since it
contains the main information between the ingredients of pointer (measuring
device) and measured system. According to the measurement theory,
the explicit expressions of Hamiltonian of the pointer and measured
system doesn't affect the results of the measurement. In our case,
the interaction Hamiltonian of a measurement is taken as the standard
von Neumann Hamiltonian 

\begin{equation}
H=g(t)\hat{\text{\ensuremath{\sigma}}}_{x}\otimes\hat{P},\label{eq:Hamil}
\end{equation}
where $g\left(t\right)$ is the interaction coupling function between
the pointer and measured system, $\hat{P}$ denotes the conjugate
momentum operator to the position operator $\hat{X}$ of the pointer
with $[\hat{X},\hat{P}]=i\hat{I}$, and $\hat{\sigma}_{x}=\vert\text{\ensuremath{\uparrow_{x}}}\rangle\langle\downarrow_{x}\vert-\vert\downarrow_{x}\rangle\langle\uparrow_{x}\vert$
is Pauli $x$ operator of the system to be measured. Here, $\vert\uparrow_{x}\rangle$
and $\vert\downarrow_{x}\rangle$ are the eigenstates of $\hat{\sigma}_{x}$
with corresponding eigenvalues $1$ and $-1$, respectively. The coupling
$g\left(t\right)$ is a nonzero function in a finite interaction time
interval $t-t_{0}$, i.e., $\int_{t_{0}}^{t}g(\tau)d\tau=g\delta\left(t-t_{0}\right)$.
For this kind of impulsive interaction, the time evolution operator
$e^{-\frac{i}{\hbar}\int Hd\tau}$ of our total system becomes as
$e^{-\frac{1}{\hbar}ig\hat{\sigma}_{x}\otimes\hat{P}}$. Hereafter,
we put $\hbar=1$ and assume all factors in $g\hat{\text{\ensuremath{\sigma}}}_{x}\otimes\hat{P}$
are dimensionless. 

In this work, we take the polarization and spatial degrees of freedom
of SPAC state as the measured system and pointer, respectively. We
assume that initially the system and pointer state are prepared in
\begin{equation}
\vert\Phi_{in}\rangle=\vert\psi_{i}\rangle\otimes\vert\phi\rangle,\label{eq:2-1}
\end{equation}
where 
\begin{equation}
\vert\psi_{i}\rangle=\cos\frac{\varphi}{2}\vert\uparrow_{z}\rangle+e^{i\delta}\sin\frac{\varphi}{2}\vert\downarrow_{z}\rangle\label{eq:3-1}
\end{equation}
and 
\begin{equation}
\vert\phi\rangle=\gamma a^{\dagger}\vert\alpha\rangle,\ \ \ \ \gamma=\frac{1}{\sqrt{1+\vert\alpha\vert^{2}}}\label{eq:4-1}
\end{equation}
are the initial states of the measured system and pointer, respectively.
Here, $\varphi\in[0,\pi]$ , $\delta\in[0,2\pi)$, $\text{\ensuremath{\alpha=re^{i\theta}}}$
is the parameter of the coherent state $\vert\alpha\rangle$, and
$\vert\phi\rangle$ is the mathematical definition of SPAC state. 

Since the operator $\hat{\sigma}_{x}$ satisfies $\hat{\sigma}_{x}^{2}=\hat{I}$,
we can write the unitary evolution operator $e^{-ig\hat{\sigma}_{x}\otimes\hat{P}}$
as 
\begin{align}
e^{-ig\hat{\sigma}_{x}\otimes\hat{P}} & =\frac{1}{2}(\hat{I}+\hat{\sigma}_{x})\otimes D\left(\frac{\Gamma}{2}\right)+\frac{1}{2}(\hat{I}-\hat{\sigma}_{x})\otimes D\left(-\frac{\Gamma}{2}\right).\label{eq:UNA1-1}
\end{align}
In the derivation of the above expression, the position operator $\hat{X}$
and momentum operator $\hat{P}$ are written in terms of the annihilation
(creation) operators $\hat{a}$($\hat{a}^{\dagger}$), of radiation
field as 
\begin{eqnarray}
\hat{X} & = & \sigma(\hat{a}^{\dagger}+\hat{a}),\label{eq:annix}\\
\hat{P} & = & \frac{i}{2\sigma}(\hat{a}^{\dagger}-\hat{a}),\label{eq:anniy}
\end{eqnarray}
where $\sigma$ is the width of the fundamental Gaussian beam. Here,
the parameter $\Gamma\equiv g/\sigma$, and $D\left(\mu\right)$ is
the displacement operator with complex $\mu$ defined by $D(\mu)=e^{\mu\hat{a}^{\dagger}-\mu^{\ast}\hat{a}}.$
Note that the coupling parameter $\Gamma$ represents measurement
strength, and the coupling between the system and pointer is defined
as weak (strong) if $\Gamma<1$$(\Gamma>1)$. We assume throughout
this work that the coupling parameter $\Gamma$ covers all the allowed
values both in weak and strong measurement regimes. 

The total initial state $\vert\Phi_{in}\rangle$ evolves by the interaction
Hamiltonian, Eq. (\ref{eq:Hamil}), as
\begin{align}
\vert\Psi\rangle & =e^{-ig\hat{\sigma}_{x}\otimes\hat{P}}\vert\psi_{i}\rangle\otimes\vert\phi\rangle.\label{eq:fi-1}
\end{align}
After taking a postselection with the state $\vert\psi_{f}\rangle=\vert\uparrow_{z}\rangle$
onto $\vert\Psi\rangle$, the final state of the pointer and its form
can obtained as 
\begin{equation}
\vert\Phi\rangle\!=\!\frac{\beta}{\sqrt{2}}\left[\left(1\!\!+\!\langle\sigma_{x}\rangle_{w}\right)D\left(\frac{\Gamma}{2}\right)+\left(1\!-\!\langle\sigma_{x}\rangle_{w}\right)D\left(\!-\frac{\Gamma}{2}\right)\right]\vert\phi\rangle,\label{eq:21}
\end{equation}
where 
\begin{align}
\beta & ^{-2}=1\!+\!\vert\langle\sigma_{x}\rangle_{w}\vert^{2}\!+\!\gamma^{2}e^{-\frac{\Gamma^{2}}{2}}\text{\ensuremath{\times}}\nonumber \\
 & \!\!Re[(1+\!\langle\sigma_{x}\rangle_{w})^{\ast}(1\!-\!\langle\sigma_{x}\rangle_{w})(\gamma^{-2}\!-\!\Gamma^{2}\!+2iIm[\alpha])e^{2\Gamma iIm\text{[}\alpha]}]\label{eq:24-1}
\end{align}
is the normalization coefficient. Here, 
\begin{equation}
\langle\sigma_{x}\rangle{}_{w}=\frac{\langle\psi_{f}\vert\hat{\sigma}_{x}\vert\psi_{i}\rangle}{\langle\psi_{f}\vert\psi_{i}\rangle}=e^{i\delta}\tan\frac{\varphi}{2}\label{eq:WV}
\end{equation}
is the weak value of the system observable $\sigma_{x}$, and $Re$
and $Im$ represents the real and imaginary parts of a complex number.
From Eq. (\ref{eq:WV}), we know that when the pre-selected state
$\vert\psi_{i}\rangle$ and the post-selected state $\vert\psi_{f}\rangle$
are almost orthogonal, the absolute value of the weak value can be
arbitrarily large. This feature is regarded as a very useful postselected
weak measurement technique which is called weak value amplification,
and bears various applications in many research fields as mentioned
in Sec.\ref{sec:2}. We have to mention that although the weak value
can take large anomalous value, it is accompanied by low successful
postselection probability $P_{s}=\vert\langle\psi_{f}\vert\psi_{i}\rangle\vert^{2}=\cos^{2}\frac{\varphi}{2}$.
In next sections we use the normalized final state $\vert\Phi\rangle$
of SPAC state after postselection to discuss the related issues.

In order to explain the weak-to-strong measurement transition in the
next section, here we introduce some concepts by following Ref. \citep{PhysRevA.103.052215}.
By taking the postselection process into account, a transition value
is introduced to represent the values of the system observable $\text{\ensuremath{\hat{\sigma}_{x}}}$
in weak and strong measurement regimes which is defined as
\begin{equation}
\sigma_{x}^{T}=\langle\Phi\vert\Psi^{\prime}\rangle,\label{eq:12-1}
\end{equation}
with 
\begin{align}
\vert\Psi^{\prime}\rangle & =\langle\psi_{f}\vert\hat{\sigma}_{x}\vert\Psi\rangle=\langle\psi_{f}\vert\psi_{i}\rangle\times\nonumber \\
 & \left[\frac{1}{2}\left(1+\langle\sigma_{x}\rangle_{w}\right)D\left(\frac{\Gamma}{2}\right)-\frac{1}{2}\left(1-\langle\sigma_{x}\rangle_{w}\right)D\left(-\frac{\Gamma}{2}\right)\right]\vert\phi\rangle.\label{eq:24}
\end{align}
 After substituting the expressions of $\vert\Phi\rangle$ and $\vert\Psi^{\prime}\rangle$
into Eq. (\ref{eq:12-1}), the value of $\sigma_{x}^{T}$ can be obtained
as 
\begin{align}
\sigma_{x}^{T} & =\frac{1}{2}\vert\beta\vert^{2}[4Re[\langle\sigma_{x}\rangle_{w}]-\gamma^{2}(1+\langle\sigma_{x}\rangle_{w})^{\ast}(1-\langle\sigma_{x}\rangle_{w})h(\Gamma)\nonumber \\
 & +\gamma^{2}(1-\langle\sigma_{x}\rangle_{w})^{\ast}(1+\langle\sigma_{x}\rangle_{w})h^{\ast}(\Gamma)],\label{eq:25}
\end{align}
with 
\begin{equation}
h(\Gamma)=e^{-\frac{\Gamma^{2}}{2}}\left(1+(\alpha^{\ast}+\Gamma)(\alpha-\Gamma)\right)e^{2\Gamma iIm[\alpha]}.\label{eq:26}
\end{equation}
It can be seen that if $\Gamma\rightarrow0$ , the transition value
reduced to the weak value $\langle\sigma_{x}\rangle_{w}$ of the observable
$\hat{\sigma}_{x}$, i.e., 
\begin{align}
\left(\sigma_{x}^{T}\right)_{\Gamma\rightarrow0} & =\langle\sigma_{x}\rangle_{w}.\label{eq:16}
\end{align}

On the other side, if $\Gamma\rightarrow\infty$, then the transition
value gives 
\begin{align}
\left(\sigma_{x}^{T}\right)_{\Gamma\rightarrow\infty} & =\frac{1}{2}\frac{\vert1+\langle\sigma_{x}\rangle_{w}\vert^{2}-\vert1-\langle\sigma_{x}\rangle_{w}\vert^{2}}{1+\vert\langle\sigma_{x}\rangle_{w}\vert^{2}}\nonumber \\
 & =\cos\delta\sin\varphi=\sigma_{x}^{c}.\label{eq:17}
\end{align}
Here, $\sigma_{x}^{c}$ is the conditional expectation value of the
system observable $\hat{\sigma}_{x}$ in strong measurement regime.
The value of $\sigma_{x}^{c}$ can be obtained by Aharonov-Bergmann-Lebowitz
rule \citep{PhysRev.134.B1410} which reads as 
\begin{align}
\sigma_{x}^{c} & =\sum_{j}a_{j}\frac{\vert\langle\psi_{f}\vert a_{j}\rangle\langle a_{j}\vert\psi_{i}\rangle\vert^{2}}{\sum_{i}\vert\langle\psi_{f}\vert a_{i}\rangle\langle a_{i}\vert\psi_{i}\rangle\vert^{2}}\nonumber \\
 & =\frac{\vert\langle\psi_{f}\vert\uparrow_{x}\rangle\langle\uparrow_{x}\vert\psi_{i}\rangle\vert^{2}-\vert\langle\psi_{f}\vert\downarrow_{x}\rangle\langle\downarrow_{x}\vert\psi_{i}\rangle\vert^{2}}{\vert\langle\psi_{f}\vert\uparrow_{x}\rangle\langle\uparrow_{x}\vert\psi_{i}\rangle\vert^{2}+\vert\langle\psi_{f}\vert\downarrow_{x}\rangle\langle\downarrow_{x}\vert\psi_{i}\rangle\vert^{2}}\nonumber \\
 & =\cos\delta\sin\varphi.\label{eq:12}
\end{align}
As we can see, the two extreme limits of the transition value $\sigma_{x}^{T}$
directly gives the corresponding values of the observable for weak
and strong measurement. Thus, the transition value can be seen as
a generalization of weak value and conditional expectation value with
considering all allowed values of corresponding system observable
through weak-to-strong measurements. 

\section{\label{sec:3}Weak-to-strong measurement transition}

The average shifts of position $x$ and momentum $p$ variables are
defined as 

\begin{equation}
\delta x=\langle\Phi\vert\hat{X}\vert\Phi\rangle-\langle\phi\vert\hat{X}\vert\phi\rangle\label{eq:6}
\end{equation}
and 
\begin{equation}
\delta p=\langle\Phi\vert\hat{P}\vert\Phi\rangle-\langle\phi\vert\hat{P}\vert\phi\rangle,
\end{equation}
respectively. Here $\text{\ensuremath{\hat{X}} }$ and $\hat{P}$
are position and momentum operators expressed in Fock space representation
in terms of the annihilation (creation) operator $\hat{a}$ ($\hat{a}^{\dagger}$)
as aforementioned in Sec. \ref{sec:2}. By using Eq. (\ref{eq:21}),
the explicit expressions of $\delta x$ and $\delta p$ can be obtained
as 

\begin{align}
\delta x & =\langle\Phi\vert\hat{X}\vert\Phi\rangle-\langle\phi\vert\hat{X}\vert\phi\rangle\nonumber \\
 & =2\sigma Re\left[\langle\Phi\vert\hat{a}\vert\Phi\rangle\right]-2\sigma Re\left[\langle\phi\vert\hat{a}\vert\phi\rangle\right]\nonumber \\
 & =\sigma\vert\beta\vert^{2}\gamma^{2}\{\vert1+\langle\sigma_{x}\rangle_{w}\vert^{2}(\Gamma\gamma^{-2}+4Re[\alpha]+2Re[\alpha]\vert\alpha\vert^{2})\nonumber \\
 & +\vert1-\langle\sigma_{x}\rangle_{w}\vert^{2}(-\Gamma\gamma^{-2}+4Re[\alpha]+2Re[\alpha]\vert\alpha\vert^{2})\nonumber \\
 & +Re[(1+\langle\sigma_{x}\rangle_{w})^{\ast}(1-\langle\sigma_{x}\rangle_{w})f(-\Gamma)]\nonumber \\
 & +Re[(1-\langle\sigma_{x}\rangle_{w})^{\ast}(1+\langle\sigma_{x}\rangle_{w})f(\Gamma)]\}\nonumber \\
 & -2\sigma\gamma^{2}(2+\vert\alpha\vert^{2})Re[\alpha],\label{eq:32-1}
\end{align}
 and 
\begin{align}
\delta p & =\langle\Phi\vert\hat{P}\vert\Phi\rangle-\langle\phi\vert\hat{P}\vert\phi\rangle\nonumber \\
 & =\frac{\hbar}{\sigma}Im\left[\langle\Phi\vert a\vert\Phi\rangle\right]-\frac{\hbar}{\sigma}Im\left[\langle\phi\vert\hat{a}\vert\phi\rangle\right]\nonumber \\
 & =\frac{\hbar}{2\sigma}\vert\beta\vert^{2}\gamma^{2}\{\vert1+\langle\sigma_{x}\rangle_{w}\vert^{2}(\Gamma\gamma^{-2}+4Im[\alpha]+2Im[\alpha]\vert\alpha\vert^{2})\nonumber \\
 & +\vert1-\langle\sigma_{x}\rangle_{w}\vert^{2}(-\Gamma\gamma^{-2}+4Im[\alpha]+2Im[\alpha]\vert\alpha\vert^{2})\nonumber \\
 & +Im[(1+\langle\sigma_{x}\rangle_{w})^{\ast}(1-\langle\sigma_{x}\rangle_{w})f(-\Gamma)]\nonumber \\
 & +Im[(1-\langle\sigma_{x}\rangle_{w})^{\ast}(1+\langle\sigma_{x}\rangle_{w})f(\Gamma)]\}\nonumber \\
 & -\frac{\hbar}{\sigma}\gamma^{2}(2+\vert\alpha\vert^{2})Im[\alpha],\label{eq:22}
\end{align}
respectively, with
\begin{align}
f(\Gamma) & =e^{-2\Gamma iIm[\alpha]}\times\nonumber \\
 & \left[2\alpha(2+\vert\alpha\vert^{2})+3\Gamma\gamma^{-2}-2\alpha^{2}\Gamma+\Gamma^{2}(\alpha^{\ast}-3\alpha)\right]e^{-\frac{\Gamma^{2}}{2}}.\label{eq:33}
\end{align}
 These average shifts are valid for all measurement regimes, and their
magnitudes can be controlled by weak value $\langle\sigma_{x}\rangle_{w}$
and coupling parameter $\Gamma$. As aforementioned, the weak-to-strong
measurement transition can occur in continuously by adjusting the
magnitude of coupling parameter $\Gamma$. If we assume the coupling
parameter is very small, then the above position and momentum shifts
are reduced to the corresponding position and momentum shifts in postselected
weak measurement regime, i.e., 

\begin{align}
\left(\delta x\right)_{\Gamma\rightarrow0} & =W_{x}=gRe\left[\langle\sigma_{x}\rangle_{w}\right]-g\frac{\partial Var(X)_{\vert\phi\rangle}}{2\sigma^{2}\partial\theta}Im\left[\langle\sigma_{x}\rangle_{w}\right]\label{eq:35}
\end{align}
and 
\begin{equation}
\left(\delta p\right)_{\Gamma\rightarrow0}=W_{p}=\frac{2g}{\hbar}Var\left(P\right)_{\vert\phi\rangle}Im\left[\langle\sigma_{x}\rangle_{w}\right],\label{eq:40}
\end{equation}
respectively. Here,

\begin{align}
Var\left(X\right)_{\vert\phi\rangle} & =\sigma^{2}\gamma^{4}\left(3+4\vert\alpha\vert^{2}\sin^{2}\theta+\vert\alpha\vert^{4}\right)\label{eq:37}
\end{align}
and 
\begin{equation}
Var(P)_{\vert\phi\rangle}=\frac{\hbar^{2}}{4\sigma^{2}}\gamma^{4}\left[3+4\vert\alpha\vert^{2}\cos^{2}\theta+\vert\alpha\vert^{4}\right]\label{eq:42-2}
\end{equation}
are the variances of position and momentum variables under the initial
SPAC state $\vert\phi\rangle$, respectively. 

\begin{figure}
\includegraphics[width=8cm]{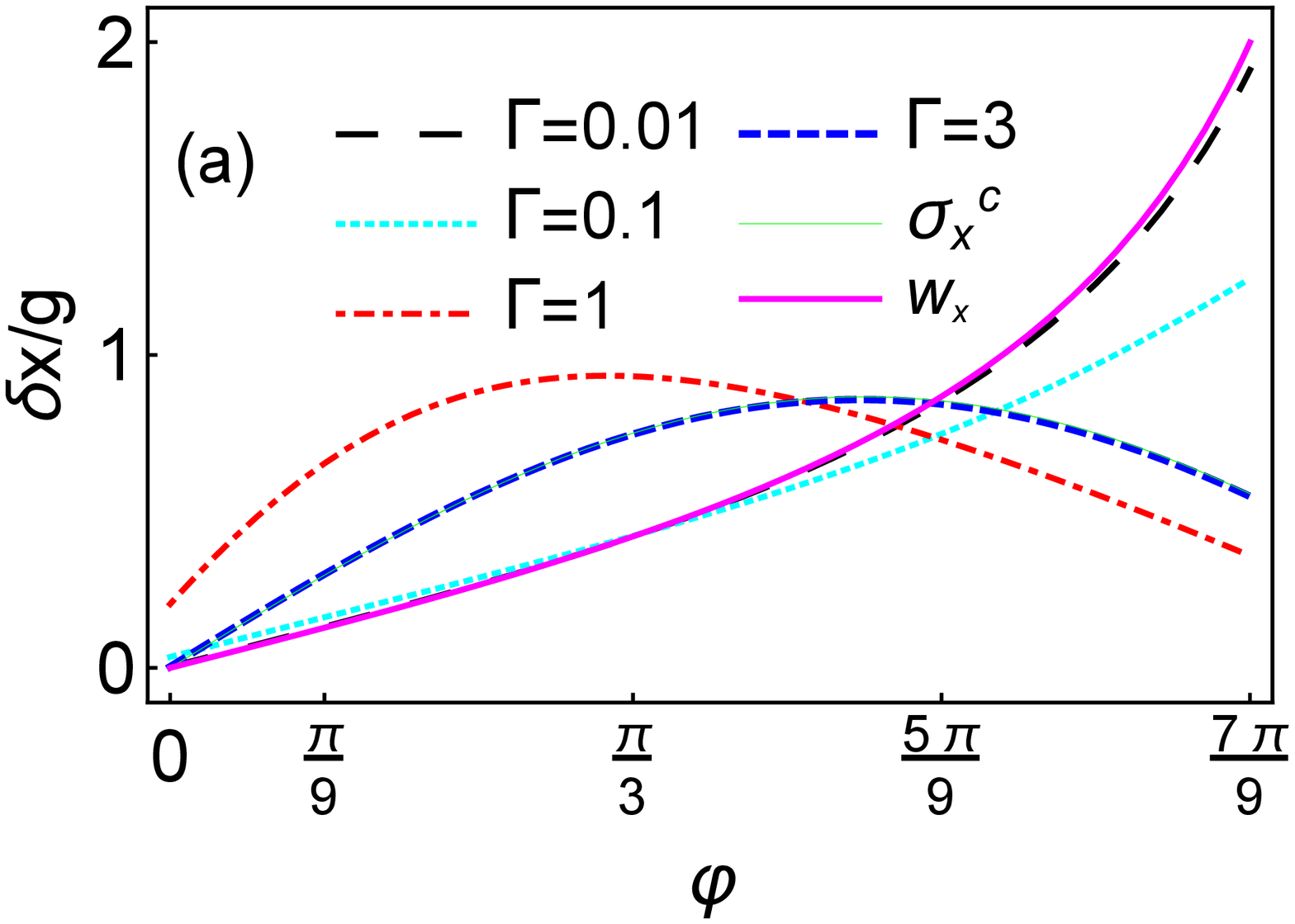}

\includegraphics[width=8cm]{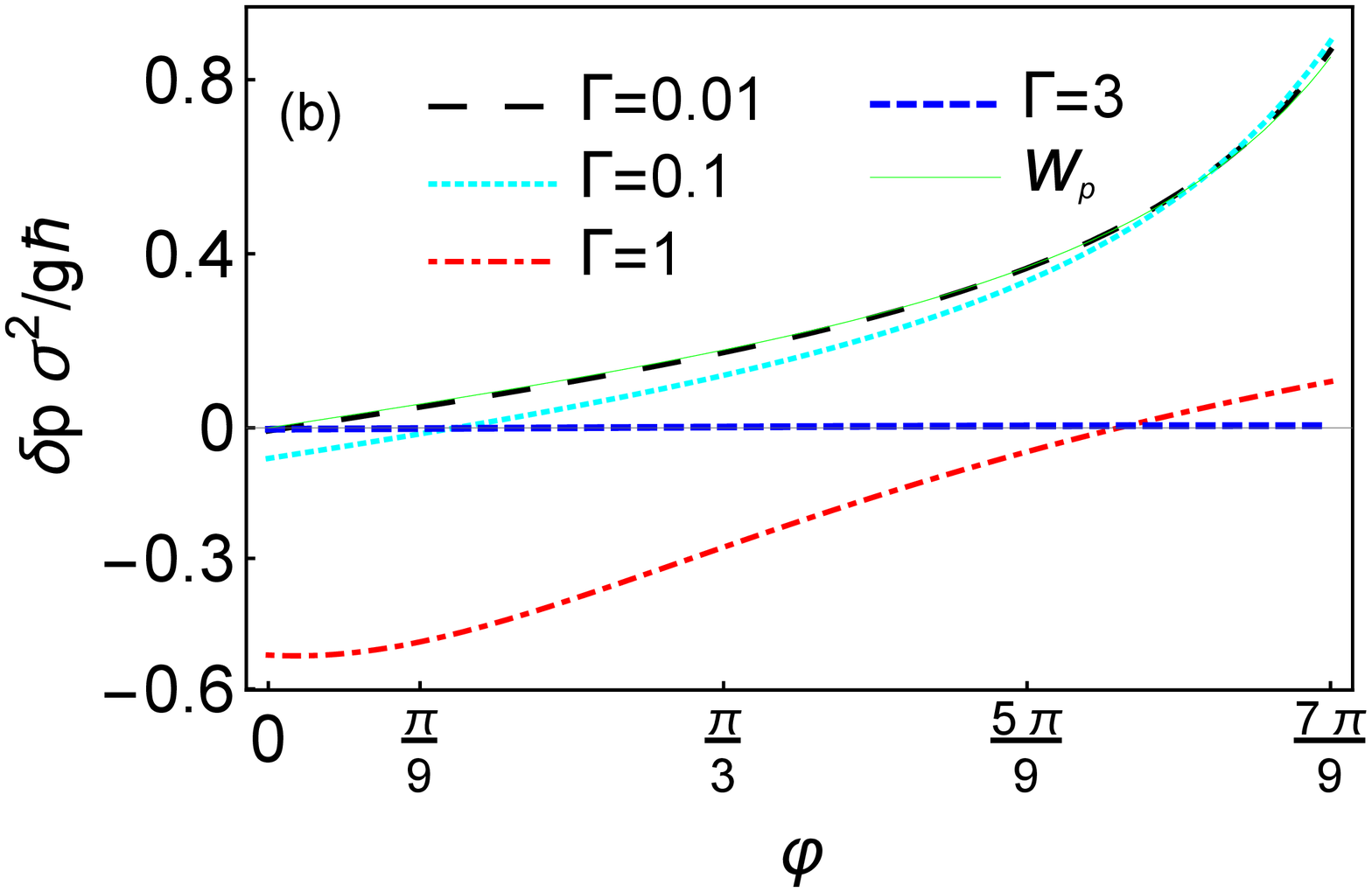}

\caption{\label{fig:1}(Color online) Measurement transition of the pointer's
shifts in the position (a) and the momentum (b) for SPAC state after
taking postselection as a function of weak value quantified by $\varphi$
for different coupling parameters $\Gamma$. Here, the other parameters
are set as $r=2,\theta=\frac{\pi}{6}$ and $\delta=\frac{\pi}{6}$.}
\end{figure}

At the other extreme, the $\delta x$ and $\delta p$ give the values
corresponding to the strong measurement regime as 

\begin{align}
\left(\delta x\right)_{\Gamma\rightarrow\infty} & =\frac{2gRe\left[\langle\sigma_{x}\rangle_{w}\right]}{1+\vert\langle\sigma_{x}\rangle\vert^{2}}=g\cos\delta\sin\varphi=g\sigma_{x}^{c},
\end{align}
 and 
\begin{equation}
\left(\delta p\right)_{\Gamma\rightarrow\infty}=0,
\end{equation}
respectively. In Fig. \ref{fig:1}, we show the weak-to-strong measurement
transition of pointer's shifts, Eq. (\ref{eq:32-1}) and Eq. (\ref{eq:22}),
as a function weak value characterized by $\varphi$. As shown in
Fig. \ref{fig:1}, the measurement transition from weak-to-strong
regime occurred in continuously with the increasing of the coupling
parameter $\Gamma$, and two extreme limits ($\Gamma\rightarrow0$
and $\Gamma\rightarrow\infty$) also match well with our theoretical
results. In the current work, the beam width $\sigma$ of SPAC state
is assumed to be fixed so that the coupling parameter $\Gamma$ is
controlled by only adjusting the coupling strength $g$ between the
pointer and measured system, which is assumed to be an impulsive interaction
independent of time. In general, the coupling strength $g$ can be
the function of time so that the measurement transition can be implemented
experimentally in three ways, each corresponding to tuning one of
the three parameters $g$, $t$ and $\sigma$. Yet, tuning of the
coupling duration time $t$ is the most straightforward approach to
implement the weak-to-strong transition \citep{Nature2020}.\textcolor{magenta}{{} }

\section{\label{sec:4}Signal-to-Noise Ratio}

In remaining parts of this work, we discuss the advantages of SPAC
state in precision measurement processes. Firstly, we investigate
the advantages of SPAC state in signal amplification process after
postselection measurement. In general, the signal amplification of
position shift can be characterized by SNR \citep{Agarwal2013}. Thus,
in order to show the advantages of postselected measurement over non-postselected
one based on SPAC state for its position shift, we investigate the
ratio of SNRs between postselected and nonpostselected measurments,
i.e., 
\begin{equation}
\chi=\frac{\mathcal{R}_{p}}{\mathcal{R}_{n}}.\label{eq:42}
\end{equation}
Here, $\mathcal{R}_{p}$ represents the SNR of postselected von Neumann
measurement defined as 

\begin{equation}
\mathcal{R}_{p}=\frac{\sqrt{NP_{s}}\delta x}{\triangle x},\label{eq:10}
\end{equation}
with the variance of position operator 
\begin{equation}
\triangle x=\sqrt{\langle\Phi\vert\hat{X}^{2}\vert\Phi\rangle-\langle\Phi\vert\hat{X}\vert\Phi\rangle^{2}}\label{eq:4}
\end{equation}
 and the average shift of the pointer variable $x$ after postselected
measurement
\begin{align}
\delta x & =\langle\Phi\vert\hat{X}\vert\Phi\rangle-\langle\phi\vert\hat{X}\vert\phi\rangle,\label{eq:45-1}
\end{align}
respectively. Here, $\hat{X}=\sigma\left(\hat{a}+\hat{a}^{\dagger}\right)$
is the position operator, $N$ is the total number of measurements,
$P_{s}$ is the success probability of postselection, and $\vert\Phi\rangle$
denotes the normalized state of SPAC state after postselection, i.e.,
Eq. (\ref{eq:21}).

\begin{figure}
\includegraphics[width=8.5cm]{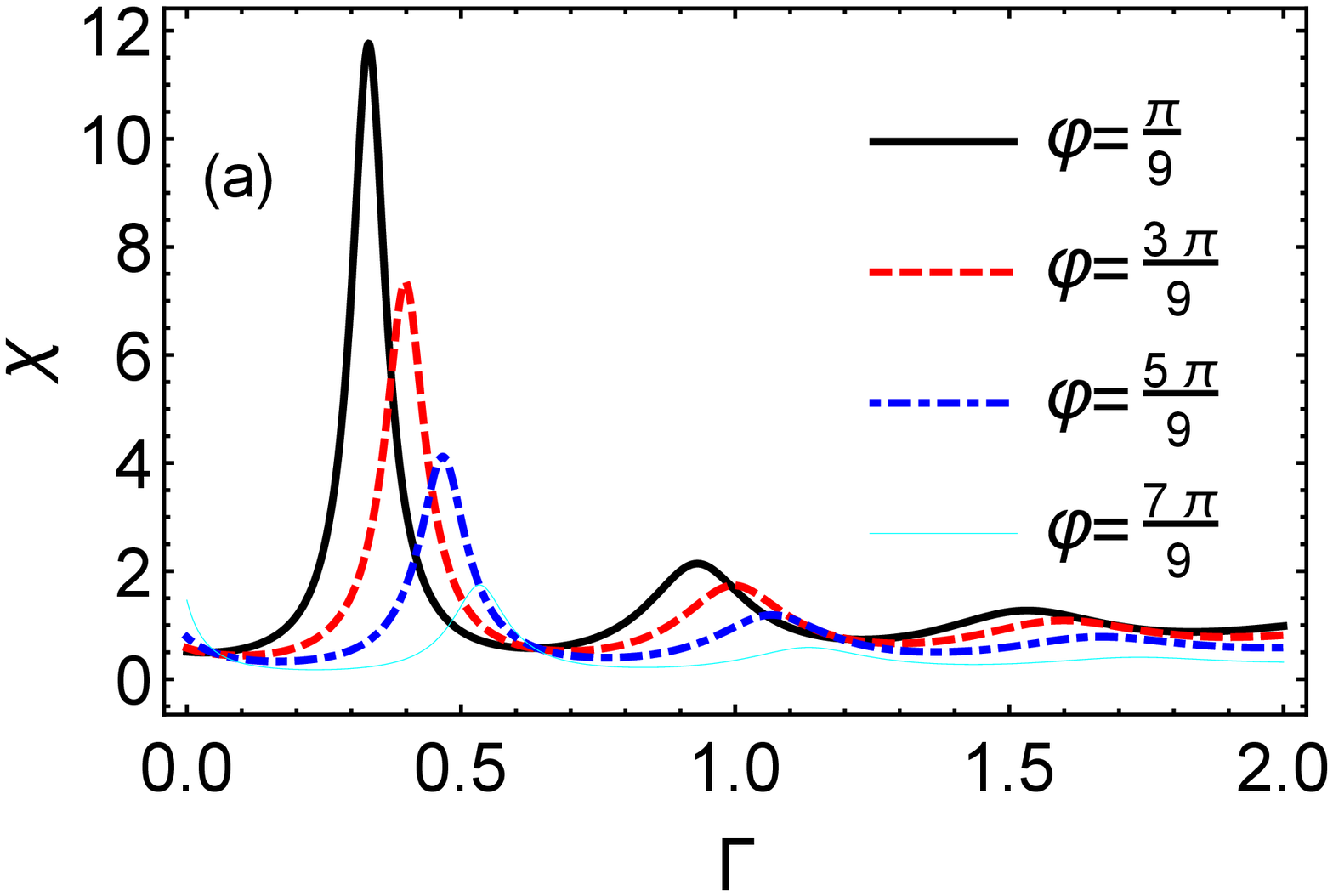}

\includegraphics[width=8cm]{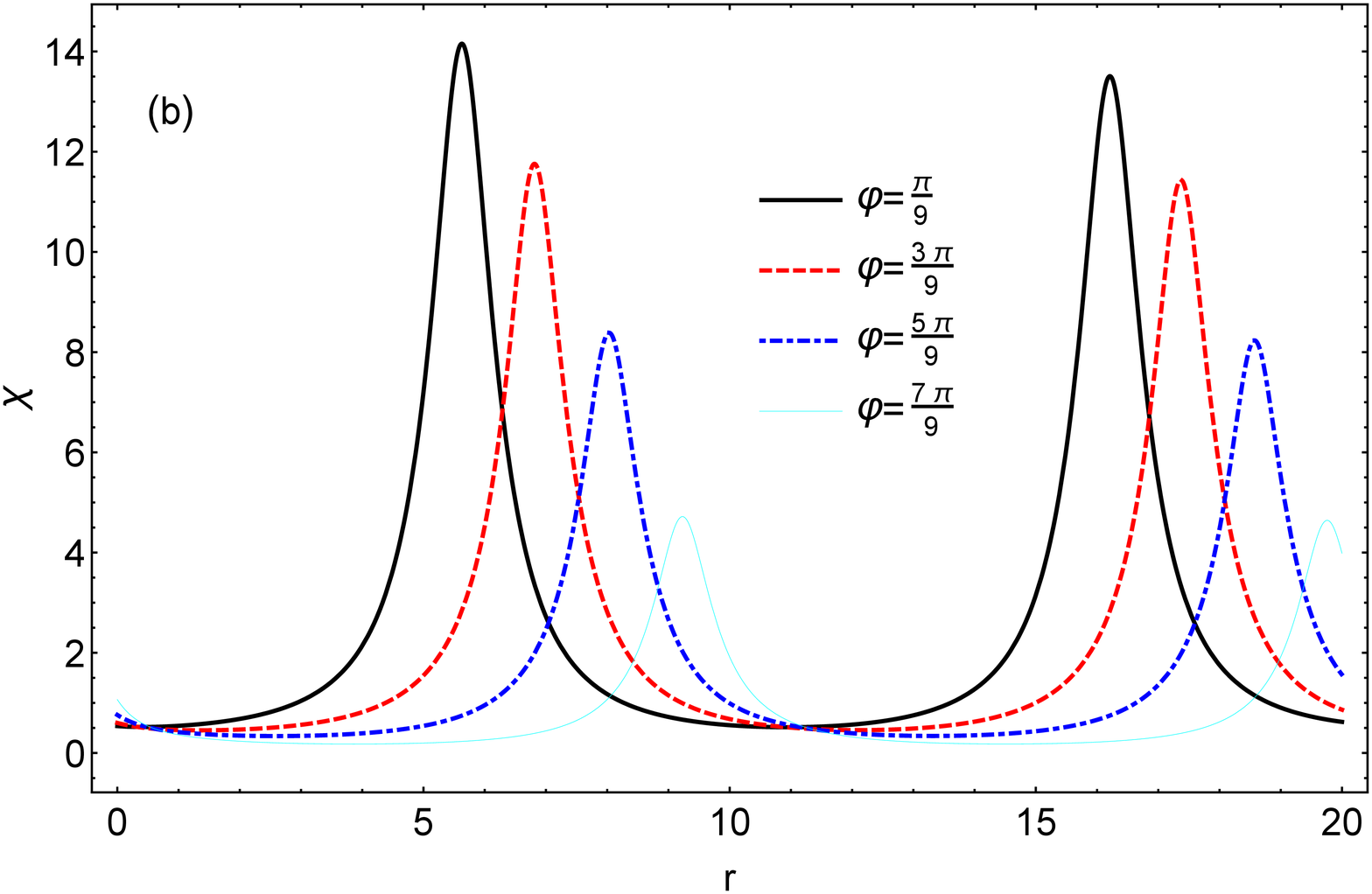}

\caption{\label{fig:3} (Color online) The ratio $\chi$ of SNRs between postselected
and nonpostselected measurement for SPAC state. Here we take $\delta=\frac{5\pi}{12},\theta=\frac{\pi}{2}$.
(a) $\chi$ is plotted as a function of coupling parameter $\Gamma$
for different weak values for fixed $r=5$. (b) $\chi$ is plotted
as a function of initial SPAC state parameter $r$ for different weak
values for the fixed coupling parameter $\Gamma=0.3$.}
\end{figure}

The $\mathcal{R}_{n}$ for non-postselected measurement is defined
as \citep{Agarwal2013} 

\begin{equation}
\mathcal{R}_{n}=\frac{\sqrt{N}\delta x^{\prime}}{\triangle x^{\prime}},\ \label{eq:11}
\end{equation}
 with

\begin{equation}
\triangle x^{\prime}=\sqrt{\langle\Psi\vert\hat{X}^{2}\vert\Psi\rangle-\langle\Psi\vert\hat{X}\vert\Psi\rangle^{2}}\label{eq:47}
\end{equation}
 and 

\begin{align}
\delta x^{\prime} & =\frac{\langle\Psi\vert\hat{X}\vert\Psi\rangle}{\langle\Psi\vert\Psi\rangle}-\langle\phi\vert\hat{X}\vert\phi\rangle=g\sin\varphi\cos\delta,\label{eq:48}
\end{align}
 respectively. Here, $\vert\Psi\rangle$ is the unnormalized final
state of the total system without postselection which is given in
Eq. (\ref{eq:fi-1}). The explicit expression of $\delta x$ is given
in Eq. (\ref{eq:32-1}) and other quantities also can be calculated
with some algebra easily. In Fig. \ref{fig:3} we show the ratio $\chi$
of SNRs between postselected and nonpostselected von Neumann measurement
for different system parameters. As presented in Fig. \ref{fig:3}
(a), the ratio $\chi$ increases with the decreasing of the weak value
in the weak measurement regime where the coupling parameters $\Gamma$
$\in(0.3,0.5)$. The ratio $\chi$ can even be much larger than unity
near $\Gamma=0.3$. Furthermore, in Fig. \ref{fig:3} (b) we plot
the ratio $\chi$ as a function of the state parameter $r$ with different
weak values for the coupling parameter $\Gamma=0.3$ corresponding
to the weak measurement regime. From the Fig. \ref{fig:3}(b) we can
see that in the weak measurement regime the ratio $\chi$ of SNRs
exhibits a slightly damping periodic oscillation with the increasing
of system parameter $r$ which characterizes the nonclassicality of
the initial SPAC state \citep{PhysRevA.43.492}. Interestingly, we
noticed that for SPAC state based postselected von Neumann measurement
the $\chi$ is decreased with very large weak values, which is contrary
to the signal amplification feature as verified in previous studies\citep{PhysRevLett.115.120401,PhysRevA.92.022109,Turek2020}.
In a word, from the above discussions we can conclude that the SPAC
state can be utilized to improve the SNR in postselected weak measurement
rather than the nonpostselected measurement. 

\section{\label{sec:5}Fisher Information }

In this section, we study the usefulness of SPAC state after postselected
weak measurement on parameter estimation process over initial input
state. Fisher information is the maximum amount of information about
the unknown parameter that we can extract from the system. For a pure
quantum state $\vert\Phi_{\Gamma}\rangle$, the quantum Fisher information
for estimating the unknown parameter $\Gamma$ is 
\begin{equation}
F=4[\langle\dot{\Phi}_{\Gamma}\vert\dot{\Phi}_{\Gamma}\rangle-\vert\langle\text{\ensuremath{\Phi}}\vert\dot{\Phi}_{\Gamma}\rangle\vert^{2}],\label{eq:45}
\end{equation}
where state $\vert\text{\ensuremath{\Phi_{\Gamma}}}\rangle$ represents
the final pointer states of the system which is given in Eq. (\ref{eq:21}),
$\vert\dot{\Phi}_{\Gamma}\rangle=\partial_{\Gamma}\vert\Phi\rangle$,
and $\Gamma\equiv g/\sigma$ is the measurement coupling parameter
which is directly related to coupling constant $g$ in our Hamiltonian,
Eq. (\ref{eq:Hamil}).

In the information theory, the Cramer-Rao bound (CRB) is the fundamental
limit in the minimum uncertainty for parameter estimation, and it
equals to the inverse of the Fisher information, i.e. 

\begin{equation}
\Delta\Gamma\geq\frac{1}{NF^{(Q)}}.\label{eq:46}
\end{equation}
Here, $N$ the total number of measurement trials, and $F^{(Q)}=P_{s}F$
is the fisher information of our scheme after taking the successful
postselection probability $P_{s}=\cos^{2}\frac{\varphi}{2}$ into
account. As indicated in Eq. (\ref{eq:46}), the larger the Fisher
information is, the better estimation of parameter $\Gamma$ achieves
for the fixed measurement trials. 

\begin{figure}
\includegraphics[width=8cm]{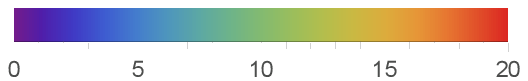}

\includegraphics{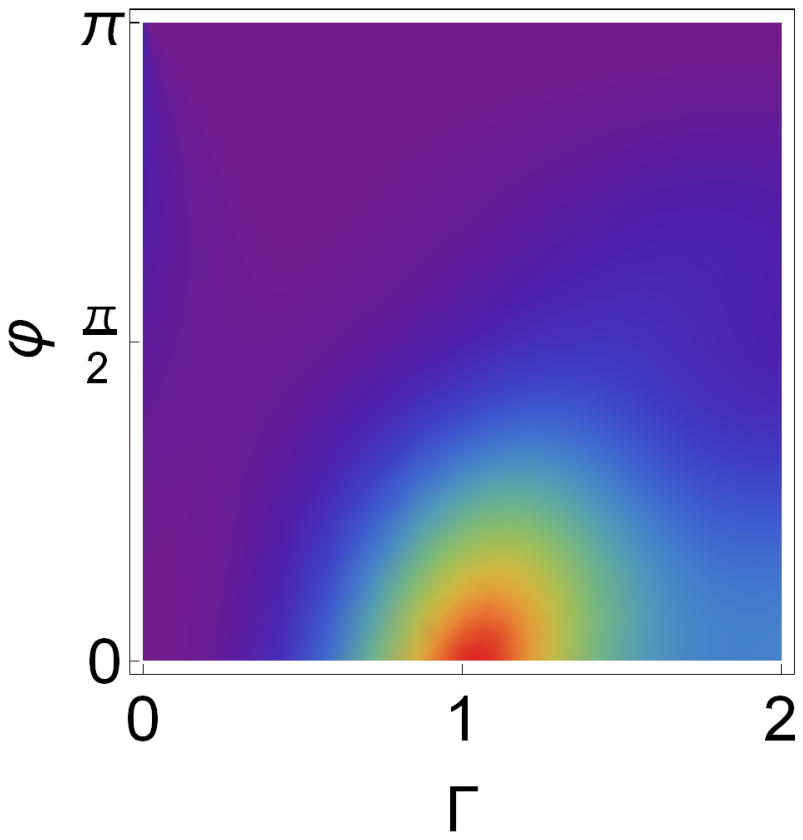}

\caption{\label{fig:4}(Color online) The quantum Fisher information $F^{(Q)}$
of SPAC state after postselected measurement as a function of coupling
parameter $\Gamma$ and weak value quantified by $\varphi$. Here,
we take $N=1$, and other parameters are the same as Fig. \ref{fig:1}.}
\end{figure}

\begin{figure}
\includegraphics[width=8cm]{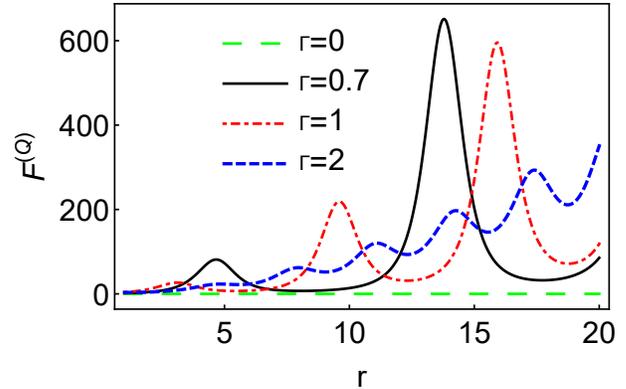}

\caption{\label{fig:5}(Color online) The quantum Fisher information $F^{(Q)}$
of SPAC state after postselected measurement as a state parameter
$r$ for different coupling parameter $\Gamma$. Here, we take $N=1$,
and other parameters are the same as Fig. \ref{fig:1}.}
\end{figure}

In order to evaluate the variance $\triangle\Gamma$, we have to calculate
the Fisher information $F^{(Q)}$ corresponding to the state $\vert\Phi\rangle$.
The analytical results for the Fisher information are presented in
Fig. \ref{fig:4} and Fig. \ref{fig:5}, respectively. In Fig. \ref{fig:4},
we plot the Fisher information $F^{(Q)}$ as a function of the coupling
parameter $\Gamma$ and weak value quantified by $\varphi$ for the
coherent state parameter $r=2$. As shown in Fig. \ref{fig:4}, the
Fisher information $F^{(Q)}$ is larger around $\Gamma=1$ with small
weak values compared to the weak measurement regime ($\Gamma\ll1$)
and nonpostselected case ($\Gamma=0$). To further verify our claims,
we plot the $F^{(Q)}$ as a function of the coherent state parameter
$r$ for different coupling parameters $\Gamma$ with small weak value,
$\langle\sigma_{x}\rangle_{w}\approx0.234+0.135i$, corresponding
to $\delta=\varphi=\frac{\pi}{6}$ in Eq. (\ref{eq:WV}). It is very
clear from the Fig. (\ref{fig:5}) that after postselected measurement
the Fisher information of SPAC state is increased significantly for
moderate coupling parameters $\Gamma$ as increasing the state parameter
$r$. 

From the discussions of weak-to-strong measurement transition in Section.
\ref{sec:3}, we can know that with increasing the coupling parameter
$\Gamma$, the weak value transformed to the conditional expectation
value. Thus, one can deduce that in strong measurement regime there
is no any amplification effect caused by the ``weak value'', and
it can be verified in numerical results presented in Figs. \ref{fig:3}-\ref{fig:5},
respectively. Furthermore, the above results also indicate that the
postselection process can improve the parameter estimation rather
than the nonpostselection one \citep{11}.

\section{\label{sec:6}conclusion and outlook}

In summary, we investigated the weak-to-strong measurement transition
and some precision measurement processes based on SPAC state after
postselected von Neumann measurement. To achieve this aim, we take
the polarization and spatial degrees of freedom of initial SPAC state
as measured system and pointer, respectively. After the standard process
of postselected weak measurement we obtained the final state of the
pointer, and discussed the related problems. Firstly, we studied the
weak-to-strong measurement transition for SPAC state via a coupling
parameter $\Gamma$ that involves the system and pointer coupling.
We found that the weak-to-strong measurement transition characterized
by pointer's shifts can occur in continuously by controlling the coupling
parameter $\Gamma$. We also investigated the advantages of SPAC state
afer postselected measurement on some precision measurement processes
such as system's signal amplification and parameter estimation. We
found that SPAC state can improve the SNR after taking postselection
in weak measurement regime for small weak values compared to nonpostselection
process. We also noticed that the ratio of SNRs between postselected
and nonpostselected von Neumann measurement can be much larger than
unity and oscillates periodically with a slight damping in the weak
measurement regime. In parameter estimation process, we calculated
the Fisher information for the final state of SPAC state after postselected
von Neumann measurement for estimating the variance of unknown coupling
parameter $\Gamma$ which is quantified by CRB. We found that SPAC
state after postselected measurement can increase the estimation of
unknown coupling parameter in moderate coupling strength regimes with
small weak value. Last but not the least, contrary to the previous
studies, the large anomalous weak values can't improve the efficiency
of related precision measurement processes based on SPAC state. 

The single photon state and the coherent state correspond to the two
limit cases (for $\vert\alpha\vert$\textrightarrow 0 or $\vert\alpha\vert$$\text{\ensuremath{\gg1}}$)
of the SPAC state. Therefore, our current postselected von Nuemman
measurement proposal based on SPAC pointer state not only covers both
of the two extreme cases, but may also provide an alternate method
to improve the related quantum information processing and quantum
metrology \citep{z2011} based on the SPAC pointer state. Another
interesting point is that, contrary to the other\textit{ }possible
pointer states including coherent state, Fock state and squeezed state
which are widely used in weak measurement studies, the SPAC state
possesses both the key features normally associated to quantum states:
the negativity of the Wigner function and the reduced fluctuations
along one quadrature. As previous works investigated, the squeezed
nature of SPAC state potentially offers an advantage in metrology
such as parameter estimation \citep{130} and quantum sensing \citep{PhysRevA.90.013821,gard2016photon,SCHNABEL20171}
compared to the coherent state and Fock state, and helps to develop
security protocols in quantum key distribution \citep{Van2006,Loepp2006,Barnett2009,PhysRevA.98.013809}.
Furthermore, in our recent work, we studied the effects of postselected
von Neumann measurement on the nonclassicality of SPAC state including
 squeezing and photon statistics \citep{yusu2021} and found that
the postselected von Neumann measurement characterized by weak value
really has positive effects to optimize the inherent properties of
SPAC state. However, the comparison of advantages of different pointer
states in postselected von Nuemann measurement based quantum metrology
is not the main goal of our present work, and it is an open problem
worth to study. Work along this line is in progress, and results will
be presented in near future. 
\begin{acknowledgments}
This work was supported by the National Natural Science Foundation
of China (Grants No. 11865017, No. 11864042), the Natural Science
Foundation of Xinjiang Uyghur Autonomous Region of China (Grant No.
2020D01A72), and the Introduction Program of High Level Talents of
Xinjiang Ministry of Science.

\end{acknowledgments}

\bibliographystyle{apsrev4-1}
\addcontentsline{toc}{section}{\refname}\bibliography{ref}

\end{document}